\documentclass[conference]{IEEEtran}
\IEEEoverridecommandlockouts
\usepackage{cite}
\usepackage{amsmath,amssymb,amsfonts}
\usepackage{algorithmic}
\usepackage{graphicx}
\usepackage{textcomp}
\usepackage{xcolor}

\usepackage{subcaption}
\usepackage{hyperref}
\hypersetup{
	pdfencoding=auto,
	linkcolor=black,
	colorlinks=true,
	citecolor=black,
	urlcolor=blue!50!black!100,
	pdftitle={ciscNet - A Single-Branch Cell Instance Segmentation and Classification Network},
	pdfauthor={Moritz Böhland, Oliver Neumann, Marcel Schilling, Markus Reischl, Ralf Mikut, Katharina Löffler, Tim Scherr},
	pdfsubject={},
	pdfkeywords={histology, cell segmentation, classification},
}
\usepackage{siunitx}
\usepackage{tabu}
\usepackage{booktabs}

\def\BibTeX{{\rm B\kern-.05em{\sc i\kern-.025em b}\kern-.08em
    T\kern-.1667em\lower.7ex\hbox{E}\kern-.125emX}}
\begin{document}

\title{ciscNet - A Single-Branch Cell Instance Segmentation and Classification Network\\
\thanks{We are grateful for funding by the Helmholtz Association in the program Natural,
Artificial and Cognitive Information Processing, HIDSS4Health - the Helmholtz Information \& Data Science School for Health and The Helmholtz Associations Initiative and Networking Fund through Helmholtz AI.\\ * contributed equally}
}
\author{\IEEEauthorblockN{1\textsuperscript{st} Moritz Böhland\textsuperscript{*}}
\IEEEauthorblockA{\textit{\footnotesize{Institute for Automation and Applied Informatics}}\\
\textit{\footnotesize{Karlsruhe Institute of Technology}}\\
\footnotesize{Eggenstein-Leopoldshafen, Germany} \\
\footnotesize{moritz.boehland@kit.edu}}
\and
\IEEEauthorblockN{2\textsuperscript{nd} Oliver Neumann}
\IEEEauthorblockA{\textit{\footnotesize{Institute for Automation and Applied Informatics}}\\
\textit{\footnotesize{Karlsruhe Institute of Technology}}\\
\footnotesize{Eggenstein-Leopoldshafen, Germany} \\
oliver.neumann@kit.edu}
\and
\IEEEauthorblockN{3\textsuperscript{rd} Marcel P. Schilling}
\IEEEauthorblockA{\textit{\footnotesize{Institute for Automation and Applied Informatics}}\\
\textit{\footnotesize{Karlsruhe Institute of Technology}}\\
\footnotesize{Eggenstein-Leopoldshafen, Germany} \\
marcel.schilling@kit.edu}
\and
\IEEEauthorblockN{4\textsuperscript{th} Markus Reischl}
\IEEEauthorblockA{\textit{\footnotesize{Institute for Automation and Applied Informatics}}\\
\textit{\footnotesize{Karlsruhe Institute of Technology}}\\
\footnotesize{Eggenstein-Leopoldshafen, Germany} \\
markus.reischl@kit.edu}
\and
\IEEEauthorblockN{5\textsuperscript{th} Ralf Mikut}
\IEEEauthorblockA{\textit{\footnotesize{Institute for Automation and Applied Informatics}}\\
\textit{\footnotesize{Karlsruhe Institute of Technology}}\\
\footnotesize{Eggenstein-Leopoldshafen, Germany} \\
ralf.mikut@kit.edu}
\and
\IEEEauthorblockN{6\textsuperscript{th} Katharina Löffler}
\IEEEauthorblockA{\textit{\footnotesize{Institute for Automation and Applied Informatics}}\\
\textit{\footnotesize{Karlsruhe Institute of Technology}}\\
\footnotesize{Eggenstein-Leopoldshafen, Germany} \\
katharina.loeffler@kit.edu}
\and
\IEEEauthorblockN{7\textsuperscript{th} Tim Scherr\textsuperscript{*}}
\IEEEauthorblockA{\textit{\footnotesize{Institute for Automation and Applied Informatics}}\\
\textit{\footnotesize{Karlsruhe Institute of Technology}}\\
\footnotesize{Eggenstein-Leopoldshafen, Germany} \\
tim.scherr@kit.edu}
}

\maketitle
\begin{abstract}
Automated cell nucleus segmentation and classification are required to assist pathologists in their decision making. The Colon Nuclei Identification and Counting Challenge 2022 (CoNIC Challenge 2022) supports the development and comparability of segmentation and classification methods for histopathological images. In this contribution, we describe our CoNIC Challenge 2022 method ciscNet to segment, classify and count cell nuclei, and report preliminary evaluation results. Our code is available at \url{https://git.scc.kit.edu/ciscnet/ciscnet-conic-2022}.
\end{abstract}


\section{Introduction}\label{sec:Intro}
Manual analysis of histopathological images is a time consuming and challenging task and a high inter-observer variability exists~\cite{graham_lizard_2021}. Interpretable computational pathology can help pathologists in their decision making, e.g., for tumor classification~\cite{Boehland2021}. Hereby, automated cell nucleus segmentation and classification, i.e., the extraction of location, cell morphology and cell type, are meaningful human-interpretable features. However, also the automated and accurate cell nucleus segmentation and classification are challenging due to domain gaps between laboratories, class imbalance, and low object resolution. The CoNIC Challenge 2022 for segmentation, classification and counting of nuclei addresses these issues~\cite{graham_conic_2021}. Many combined segmentation and classification methods use either two separate models or multiple branches for segmentation and classification~\cite{MoNuSAC2020, GRAHAM2019}. This increases training time and model complexity. Here, we propose a simple single-branch convolutional neural network (CNN) that learns cell instance segmentation and classification jointly.

\section{Data}\label{sec:Data}
We use the patch-level lizard dataset from the CoNIC Challenge \cite{graham_lizard_2021, graham_conic_2021}, consisting of 4981 Haematoxylin and Eosin stained histology image patches of size 256\,px$\times$256\,px. The ground truth (GT) provides nuclei instance segmentation and semantic segmentation of the nuclei classes: epithelial, lymphocyte, plasma, eosinophil, neutrophil, or connective tissue.

\section{Method}\label{sec:Method}
\subsection{Training data representation}
\begin{figure*}
    \centering
  \begin{subfigure}{0.14\textwidth}
    \includegraphics[width=\textwidth]{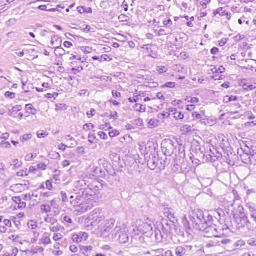}
    \setlength{\abovecaptionskip}{-8pt} 
    \caption{Image}\label{fig:DataRep_Image} 
  \end{subfigure}
  \!
  \begin{subfigure}{0.14\textwidth}
    \includegraphics[width=\textwidth]{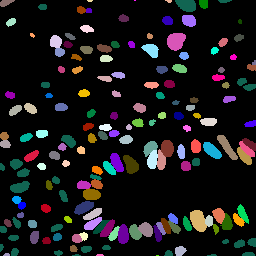}
    \setlength{\abovecaptionskip}{-8pt} 
    \caption{Instance GT}\label{fig:DataRep_Instance} 
  \end{subfigure}
  \!
  \begin{subfigure}{0.14\textwidth}
    \includegraphics[width=\textwidth]{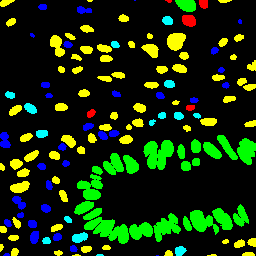}
    \setlength{\abovecaptionskip}{-8pt} 
    \caption{Semantic GT}\label{fig:DataRep_Semantic} 
  \end{subfigure}
  \!
  \begin{subfigure}{0.14\textwidth}
    \includegraphics[width=\textwidth]{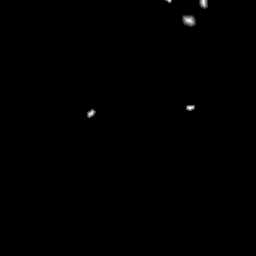}
    \setlength{\abovecaptionskip}{-8pt} 
    \caption{Neutrophil DM}\label{fig:DataRep_C1} 
  \end{subfigure}
  \!
  \begin{subfigure}{0.14\textwidth}
    \captionsetup[figure]{font=Large,labelfont=Large}
    \includegraphics[width=\textwidth]{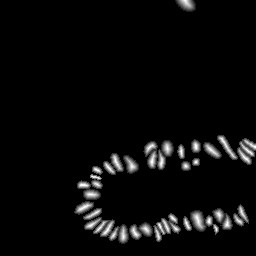}
    \setlength{\abovecaptionskip}{-8pt} 
    \caption{Epithelial DM}\label{fig:DataRep_C2} 
  \end{subfigure}
  \
  
  \smallskip
  
  \begin{subfigure}{0.14\textwidth}
    \includegraphics[width=\textwidth]{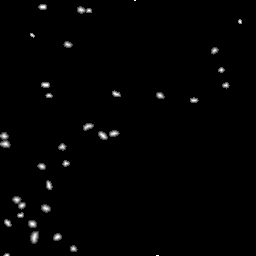}
    \setlength{\abovecaptionskip}{-8pt} 
    \caption{Lymphocyte DM}\label{fig:DataRep_C3} 
  \end{subfigure}
  \!
  \begin{subfigure}{0.14\textwidth}
    \includegraphics[width=\textwidth]{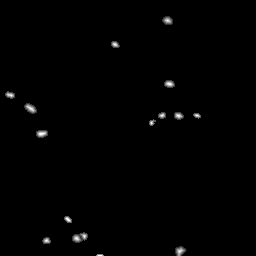}
    \setlength{\abovecaptionskip}{-8pt} 
    \caption{Plasma DM}\label{fig:DataRep_C4} 
  \end{subfigure}
  \!
  \begin{subfigure}{0.14\textwidth}
    \includegraphics[width=\textwidth]{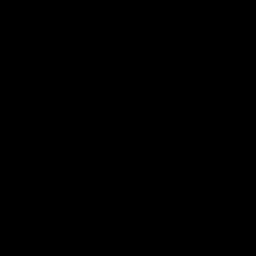}
    \setlength{\abovecaptionskip}{-8pt} 
    \caption{Eosinophil DM}\label{fig:DataRep_C5} 
  \end{subfigure}
  \!
  \begin{subfigure}{0.14\textwidth}
    \includegraphics[width=\textwidth]{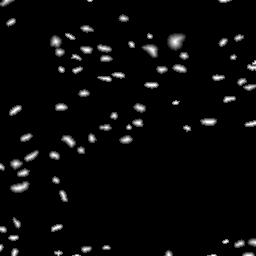}
    \setlength{\abovecaptionskip}{-8pt} 
    \caption{Connective DM}\label{fig:DataRep_C6} 
  \end{subfigure}
  \!
  \begin{subfigure}{0.14\textwidth}
    \includegraphics[width=\textwidth]{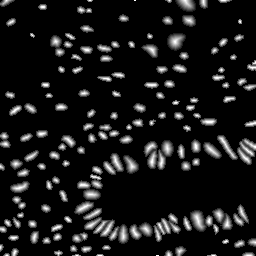}
    \setlength{\abovecaptionskip}{-8pt} 
    \caption{Overall DM}\label{fig:DataRep_Distance} 
  \end{subfigure}
  \caption{Training data representation. For nuclei of each class, normalized distance maps (DM) are computed which allow distinguishing single instances. The DMs of the single classes (d-i) sum up to the overall DM (j).}
  \label{fig:DataRep} 
\end{figure*}

To distinguish single cell instances, we use per cell normalized Euclidean distance maps. The distance maps are created from the GTs for each class separately (cf. Fig. \ref{fig:DataRep}).

\subsection{CNN architecture}\label{sec:Network}
\begin{figure*}
    \centering
    \includegraphics{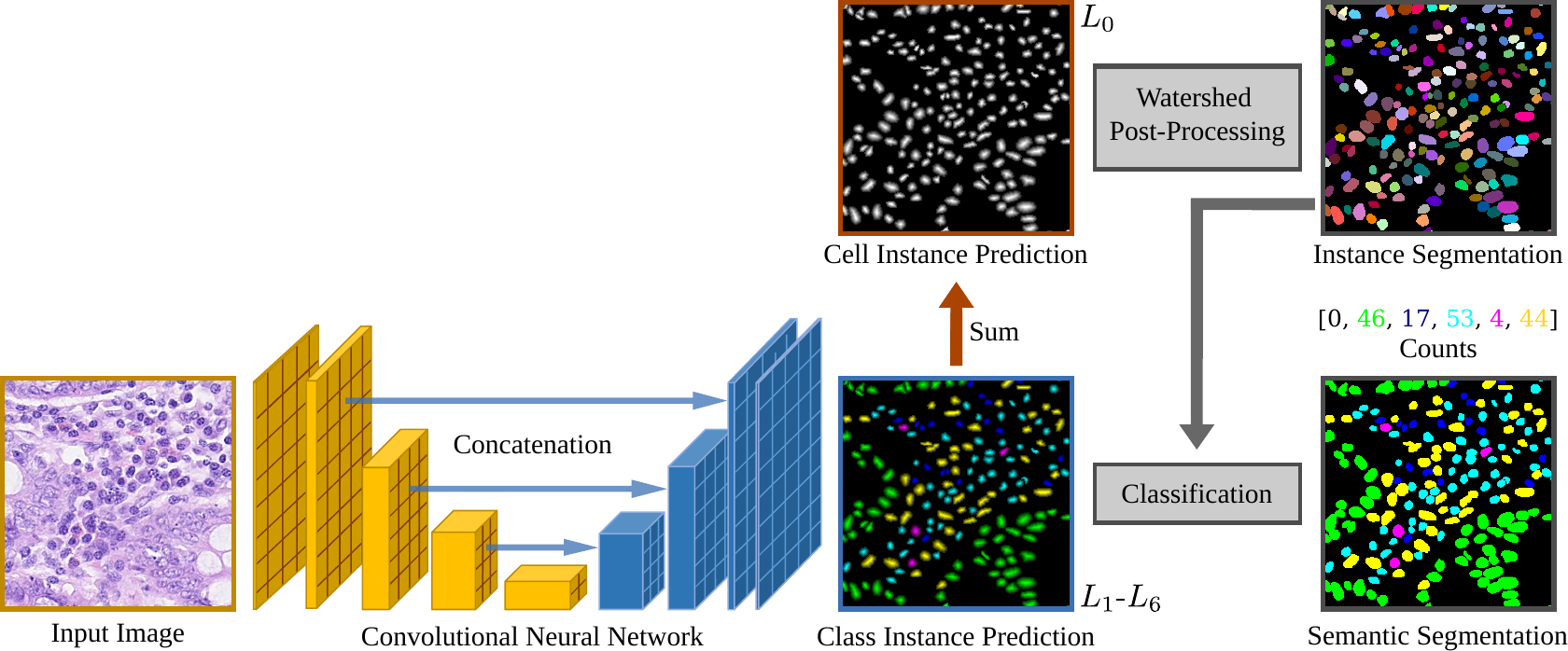}
    \caption{Method overview. The CNN is trained to predict the six cell classes (class instance prediction, losses $L_1$ - $L_6$). In addition, the sum of the six prediction channels is computed (loss $L_0$). The watershed post-processing yields the instance segmentation which is used together with the class instance prediction for classification and counting. For visualization purposes, each class has been assigned a unique color.}
    \label{fig:method-overview_2}
\end{figure*}

We use a single-branch U-Net architecture~\cite{ronneberger2015} with group normalization, 2D convolutions with stride 2 for downsampling and transposed convolutions for upsampling. The number of feature maps ranges from 64 to 1024. The mish activation function~\cite{misra2019mish} is used within the network and a linear activation for the output layer. Fig.~\ref{fig:method-overview_2} provides an overview of the architecture. The output layer predicts six classes corresponding to the six cell types (class instance prediction). In addition, the sum of the six classes is computed (cell instance prediction).

\subsection{Training data and pre-processing}
140 patches without nuclei are removed and the remaining patches are split randomly into a training set and a validation set with a ratio of 80\% to 20\%. The images are normalized from the range $[0, 255]$ to $[-1, 1]$.


\subsection{Training}\label{sec:Training}
Models are trained using the Ranger optimizer~\cite{Ranger} with a learning rate scheduler and subsequent cosine annealing. We apply the following augmentations randomly: (i) flipping/rotation, (ii) hue, saturation \& contrast variations, (iii) scaling, (iv) Gaussian blur, and (v)~Gaussian noise.

To learn the class instance prediction, the smooth L1 loss of each channel ($L_1$ - $L_6$, cf. Fig.~\ref{fig:method-overview_2}) is used. As auxiliary loss, the smooth L1 loss of the cell instance prediction ($L_0$), which is the sum of the class instance prediction channels, is used. Weight maps with increased weights for nuclei regions are applied and the single loss terms are weighted with weights $\omega_i$ to tackle the class imbalance. This yields the final overall loss $L = \sum_{i=0}^{6} \omega_i \, L_i$.


\subsection{Post-processing}\label{sec:Post-processing}
We use a watershed post-processing to obtain the instance segmentation from the cell instance prediction. Two thresholds are required: (i) a cell size threshold, and (ii) a seed extraction threshold. 
For each segmented cell, the class instance prediction is used for classification, i.e., the class instance predictions are summed up within a predicted cell and the class with the highest sum is selected. Finally, the cells of each class are counted.

\section{Results}\label{sec:Results}
\begin{table}[t]
	\center
	\footnotesize
	\caption{Results on the validation and preliminary test set with and without test-time augmentation (tta). For the CoNIC Challenge metrics mPQ\textsuperscript{+} and R2 refer to~\cite{graham_conic_2021}.}
	\begin{tabu} to 1 \columnwidth {X[5,l] X[3,c] X[4,c] X[3,c] X[3,c]}
	\toprule
	Evaluation set & mPQ\textsuperscript{+} & mPQ\textsuperscript{+} (tta) & R2 & R2 (tta)\\
	\midrule
	val & \num{0.5773} &  \num{0.5912} & \num{0.9218} & \num{0.9248}\\
	preliminary test & \num{0.4299} &  \num{0.4299} & \num{0.5587} & \num{0.5632}\\
	\bottomrule
	\end{tabu}
	\label{tab:results}
\end{table}

Table~\ref{tab:results} shows the results of our final CoNIC Challenge model on our validation set and on the preliminary challenge test set, which is a challenging subset of the final test set. For the final results on the test set (with test-time augmentation), please refer to the official challenge results.


\end{document}